\documentclass[a4paper,11pt]{article}
\usepackage{pos}

\title{Search for new phenomena in leptonic final states at CMS}

\author*[a ]{Saranya Samik Ghosh for the CMS Collaboration}
\renewcommand{\printHeadAuthors}{Saranya Samik Ghosh}

\affiliation[a]{RWTH Aachen University,\\
  III. Physikalisches Institut A., Otto-Blumenthal-Str. 16, 52074 Aachen, Germany}

\emailAdd{saranya.ghosh@cern.ch}

\abstract{Many new physics models, e.g., compositeness, extra dimensions, extended Higgs sectors, supersymmetric theories, and dark sector extensions, are expected to manifest themselves in the final states with leptons. Searches in CMS for new phenomena in the final states that include leptons, focusing on the recent results obtained using the Run-II data set collected at the LHC are described here.}

\FullConference{%
  40th International Conference on High Energy physics - ICHEP2020\\
  July 28 - August 6, 2020\\
  Prague, Czech Republic (virtual meeting)
}

\begin{document}
\maketitle

\section{Introduction}The Compact Muon Solenoid (CMS) experiment \cite{bib:CMS} at the CERN Large Hadron Collider (LHC) has a wide ranging physics program covering, but not limited to, studies of the Higgs boson, measurements of the standard model (SM), flavour physics, and a large program for searching for physics beyond the standard model (BSM). 

The CMS experiment continues with its physics program at the second run of the LHC (Run II) with proton-proton (pp) collisions at center-of-mass energy of 13 TeV from 2015 through to 2018. The increase in the center-of-mass energy provides access to wider regions in the phase space for searches for new physics and search analyses also benefit from a significantly larger dataset collected at 13 TeV, corresponding to close to 140 $\mathrm{fb^{-1}}$. 

This document presents recent results from physics analyses searching for exotic new physics phenomena in leptonic final states.  Many new physics models, e.g., compositeness, extra dimensions, extended Higgs sectors, supersymmetric theories, and dark sector extensions, are expected to manifest themselves in the final states with leptons. This is only a small subset of the physics analyses at CMS, a complete list of publications of the CMS collaboration can be found in Ref.~\cite{bib:cms_public_web}.

\section{Recent physics results on searches for exotic new physics phenomena from CMS}
\label{sec:physics_results}

Selected recent physics results regarding the searches for exotic new physics phenomena in leptonic final states are presented in the following sections.


\subsection{High mass resonant phenomena in dilepton final states}
\label{subsec:highmass}

\begin{figure}[htbp]
\begin{center}
\includegraphics[width=14cm,clip]{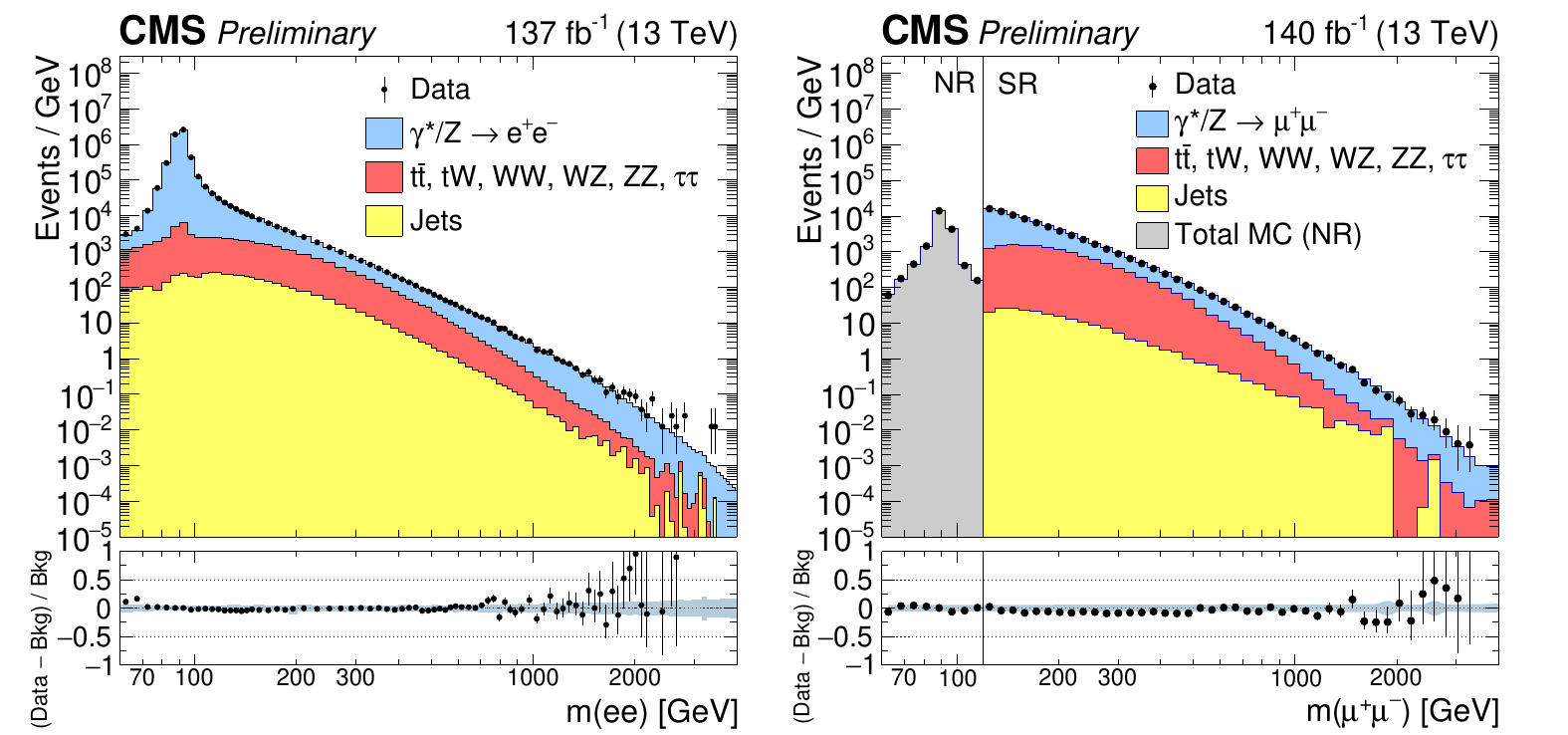}
\caption{The invariant mass distribution of pairs of (left) electrons and (right) muons observed in data (black dots with statistical error bars) and expected from the SM processes (stacked histograms). The bin width gradually increases with mass  \cite{bib:pas_cms_exo_19_019}. The ratio of the data yields after background subtraction to the background yields is shown on the bottom plots. The blue band represents the various statistical and systematic uncertainties on the background.}
\label{fig:dilep_invmass}       
\end{center}
\end{figure}

Search for high mass resonance is conducted for the case where the resonance decays into  electron or muon pairs \cite{bib:pas_cms_exo_19_019}. The analysis searches for deviations from the SM predictions in the dilepton invariant mass spectrum for both dielectron and dimuon events. 
Figure~\ref{fig:dilep_invmass}  shows the distributions of the dielectron and dimuon  invariant mass in data and the estimated background. The observed dilepton invariant mass spectrum is well described by the background expectation, and no significant evidence for the production of new particles is observed. 

Since no significant deviation from SM expectation is observed, limits are set in context of a sequential standard model for a heavy $Z'_{SSM}$ boson, and also in terms of a superstring-inspired model for $Z'_{\psi}$ particle, with the lower mass limits set at 95\% confidence level (CL)  at 5.15 TeV  and 4.56 TeV for the two respectively.

\subsection{Low mass resonance search}
\label{subsec:lowmass}

Searches have been conducted for resonances in the lower mass range for which special data taking technique, known as data scouting, is employed using special triggers and storing significantly smaller amount of event information.

A search has been performed for a narrow resonance decaying to a pair of muons in the 11.5-75 and 110-200 GeV resonance mass ranges \cite{bib:cms_lowmass_dimuon}. The search in the 45-75 and 110-200 GeV resonance mass ranges uses fully reconstructed event data, whereas the search in the 11.5-45.0 GeV mass range data collected using a dedicated set of high rate dimuon triggers with low thresholds that store a reduced amount of information. The data set analysed based on the fully reconstructed event data corresponds to an integrated luminosity of 137 $\mathrm{fb^{-1}}$ while the data set analysed using the scouting muon triggers corresponds to a total integrated luminosity of 96.6 $\mathrm{fb^{-1}}$.

\begin{figure}[htbp]
\begin{center}
\includegraphics[width=9cm,clip]{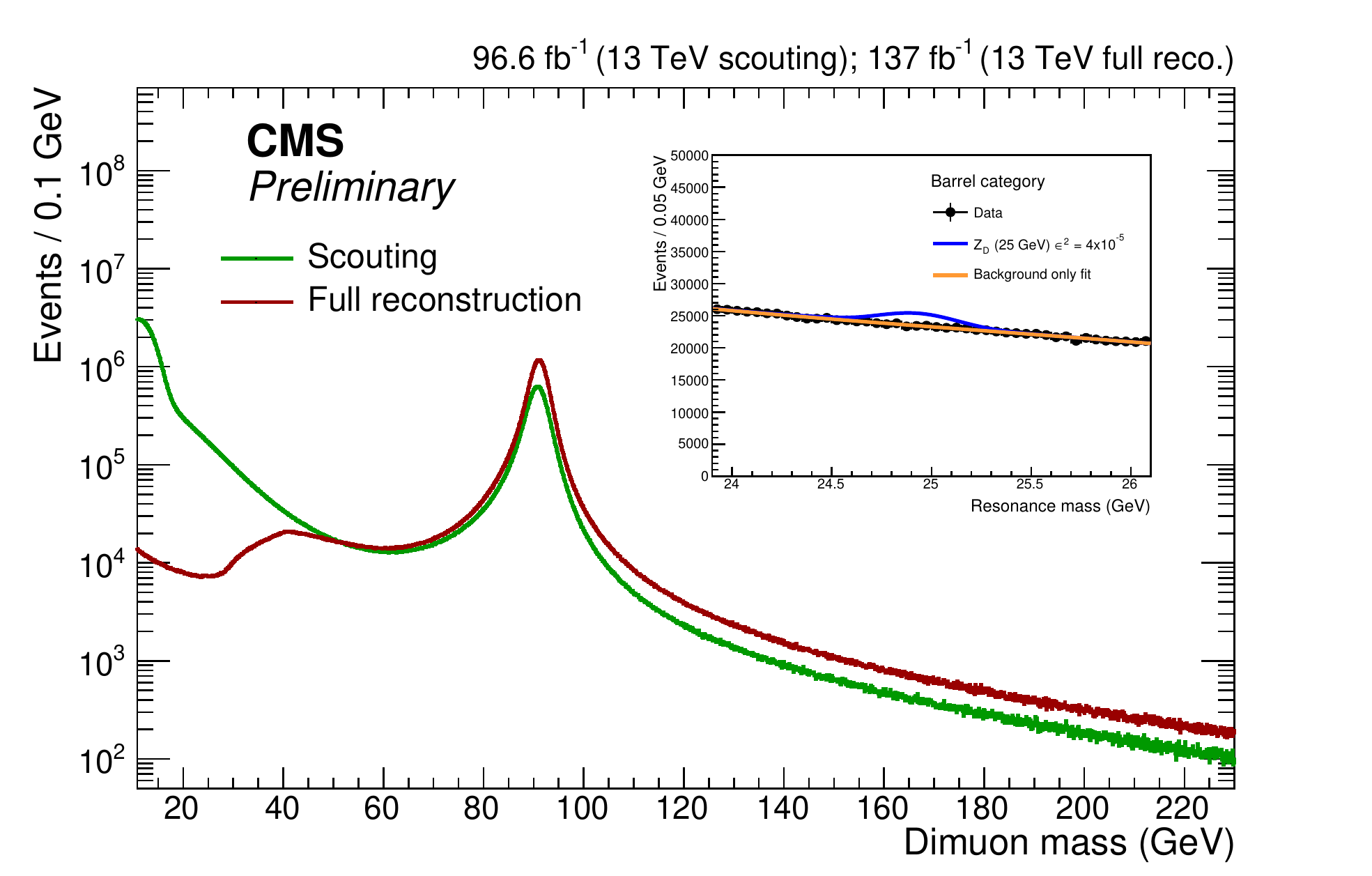}
\caption{Dimuon invariant mass distributions of events selected with the standard and scouting dimuon triggers \cite{bib:cms_lowmass_dimuon}. The distribution shown in red corresponds to the data collected using the standard triggers, corresponding to a total integrated luminosity of 137 $\mathrm{fb^{-1}}$. The distribution shown in green corresponds to the data collected using the scouting muon triggers, corresponding to a total integrated luminosity of 96.6 $\mathrm{fb^{-1}}$. The inset figure shows the dimuon mass distribution of events in the barrel category in the mass range 23.9-26.1 GeV. A function describing the background is fit to this data and a 25 GeV dark photon signal is added.}
\label{fig:lowm_dilep_invmass}       
\end{center}
\end{figure}

The dimuon invariant mass distribution, shown in Figure~\ref{fig:lowm_dilep_invmass}, is scanned and no significant resonant peaks are observed. The search sets the strongest constraints on a hypothetical dark photon heavier than 11.5 GeV.

\subsection{Search in multilepton final states}
\label{subsec:multilep}

Search for new physics phenomena has been conducted in events with three or more electrons or muons \cite{bib:cms_multilepton}. This search is sensitive to Type-III seesaw heavy fermions in the form of non-resonant excesses in the tails of the transverse mass, and to the production of light scalar or pseudoscalar boson with a pair of top quarks in the form of a resonance in the dilepton mass spectra.

The analysed data set corresponds to an integrated luminosity of 137 $\mathrm{fb^{-1}}$. Events with three or more leptons (electrons or muons) are selected and the events are classified to different signal regions based on lepton flavour, charge and invariant mass. The background is estimated using a combination of simulation and data driven methods.

No significant excess has been found, and exclusion limits set at 95\% CL on heavy fermions of the type-III seesaw model at mass of 880 GeV, and new scalar (pseudoscalar) bosons with coupling of 0.003 (0.03) and 0.04 (0.03) are excluded for masses ranges of 15-75 GeV and 108-340 GeV.

\subsection{Search for excited leptons}
\label{subsec:excitedlep}

\begin{figure}[htbp]
\begin{center}
\includegraphics[width=14cm,clip]{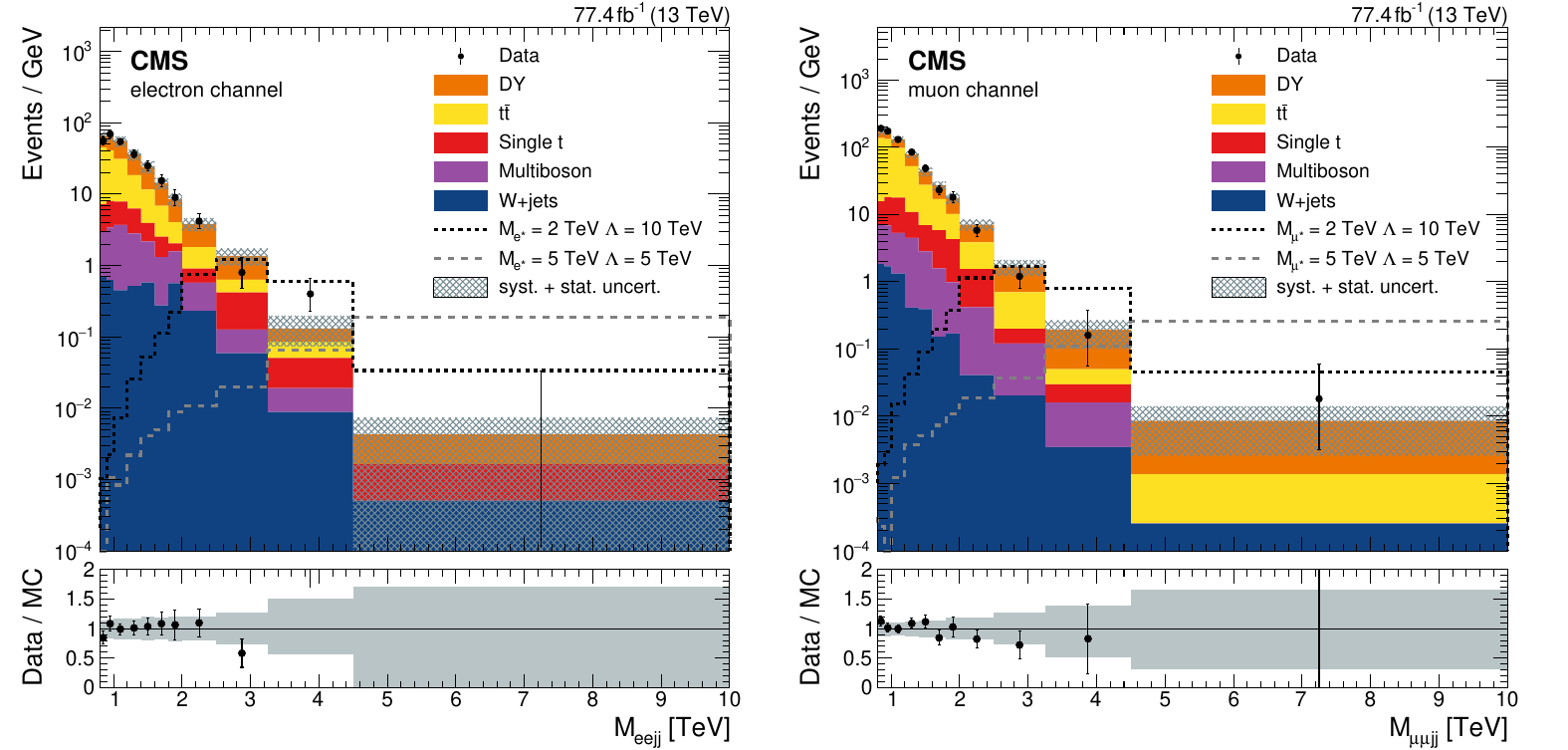}
\caption{Distribution of the two lepton two jet invariant mass in the signal region for the electron (left) and muon (right) channels \cite{bib:cms_excitedlepton}. The example signal shape for two excited lepton masses is indicated as a grey line with the parameters given in the legend. The panel below shows the data/MC ratio with the total uncertainty in grey.}
\label{fig:exlep}       
\end{center}
\end{figure}

A search for excited leptons decaying via contact interaction to final states of two electrons or two muons and two resolved jets has been carried out. This channel is most sensitive to very heavy excited leptons. The utilised data set for this analysis was recorded with the CMS detector in the years 2016 and 2017, corresponding to a total integrated luminosity of 77.4 $\mathrm{fb^{-1}}$ of proton-proton collisions at a center-of-mass energy of 13 TeV.

Di-muon and di-electron events with 2 jets are selected, and a search for signal is conducted in the 4 object invariant mass distribution following background estimation using simulation, as shown in Figure~\ref{fig:exlep}.

No significant deviations were observed in the signal region and 95\% exclusion limits are set. Excited electrons (muons) up to masses of $\mathrm{M_{e^{*}}=5.6}$ TeV ($\mathrm{M_{\mu^{*}}= 5.7}$ TeV are excluded with the usual assumption of $\mathrm{M_{\ell^{*}}= \Lambda}$, where $\Lambda$ is the substructure scale.  These are the best limits to date. The limit was also re-evaluated in terms of the substructure scale$\Lambda$ leading to a limit of $\Lambda$ = 13 TeV for masses around 2 TeV.

\section{MUSiC : Model Unspecific Search in CMS}
\label{sec:music}

MUSiC, Model Unspecific Search in CMS, is a general model independent search for deviations from the standard model for potential signs of new physics phenomena. 
The MUSiC algorithm classifies events into hundreds of different final states based on event topology, taking into account the reconstructed physics objects that are electrons, muons, photons, jets, b-jets and missing transverse momentum, and then scans each final state to search for deviations from SM expectation. The SM expectation is obtained from simulation of the different SM processes.

\begin{figure}[hp]
\begin{center}
\includegraphics[width=7cm,clip]{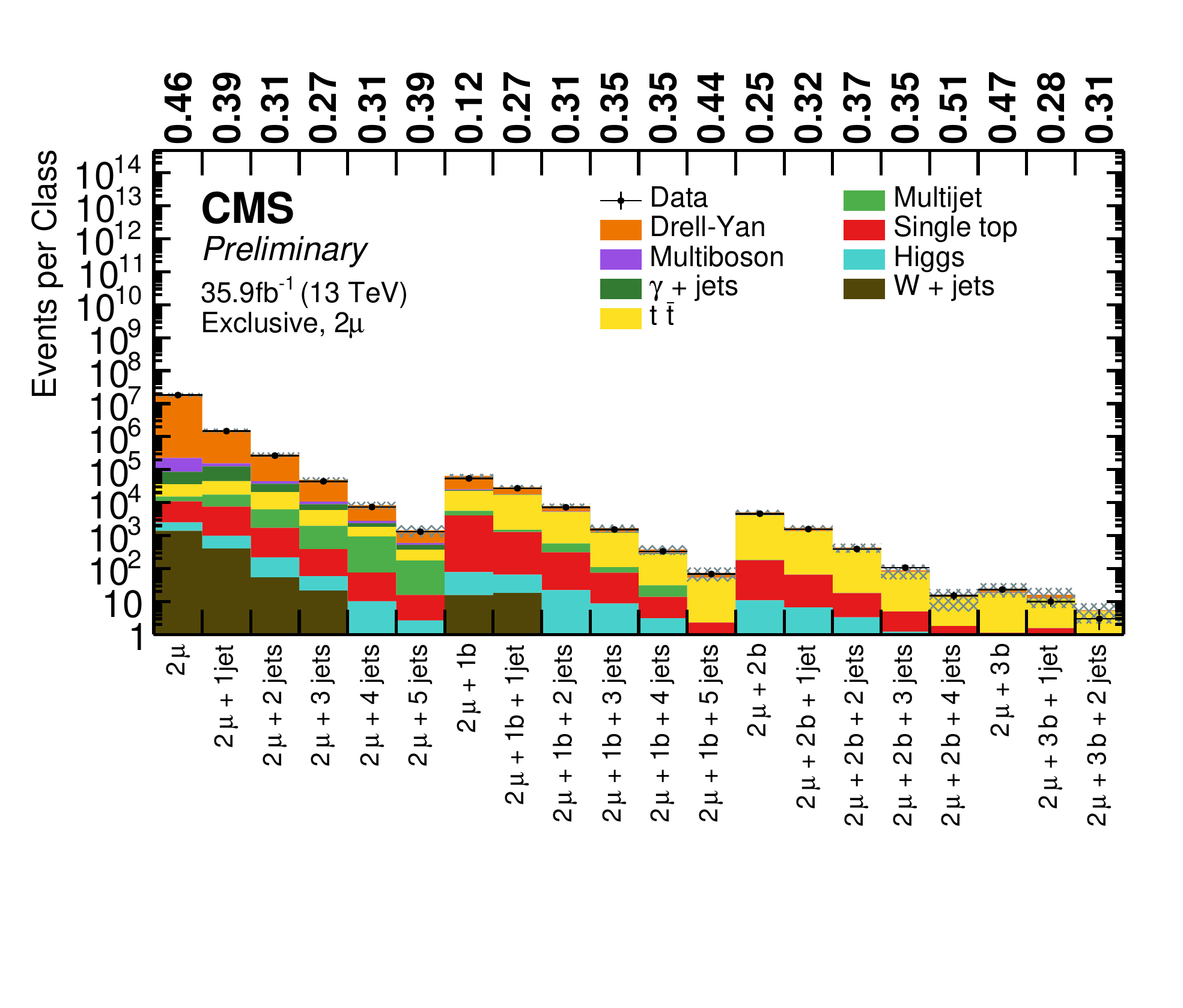}
\includegraphics[width=7cm,clip]{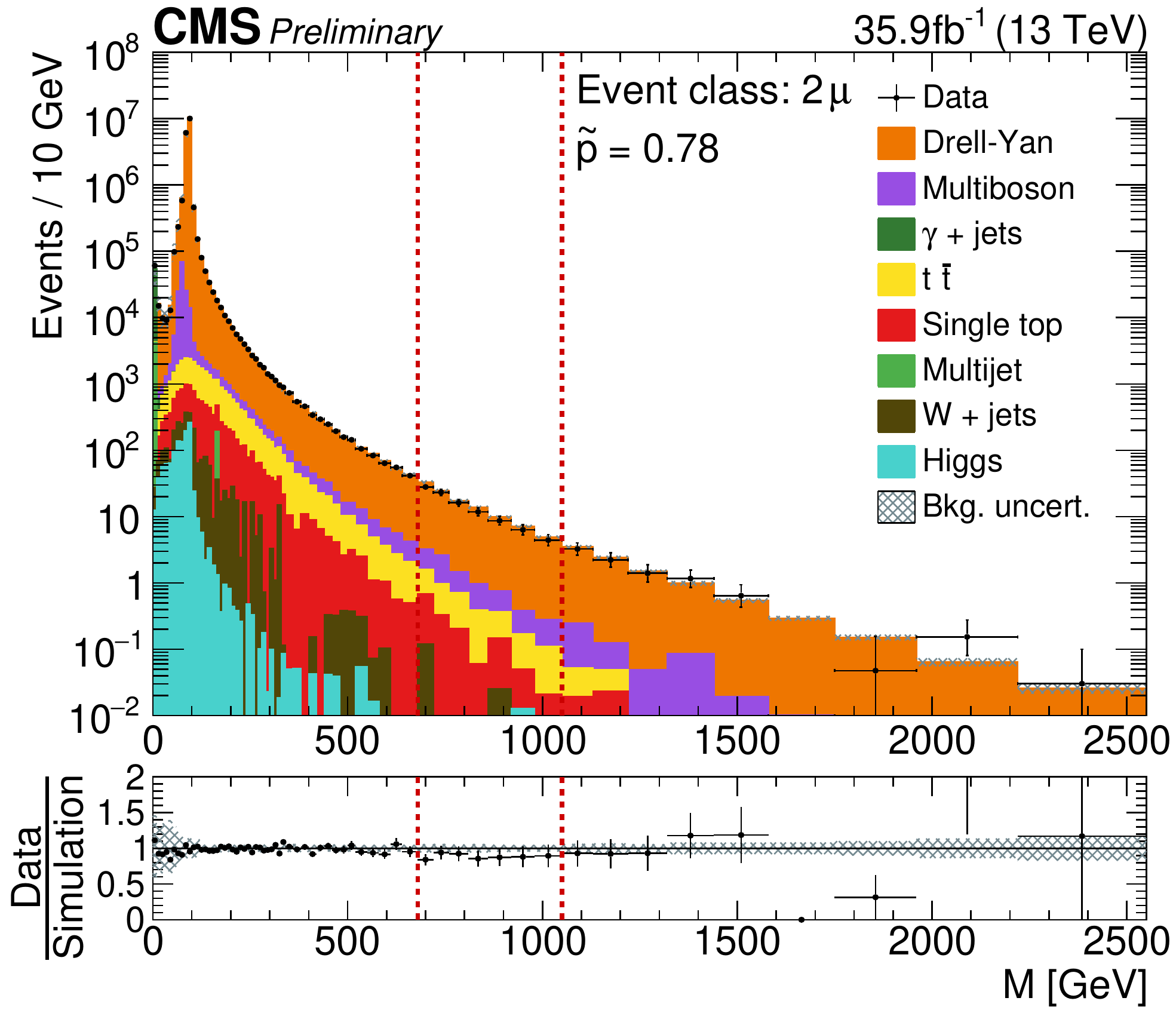}
\caption{The plot on the left shows an overview of total contributions (single bin) for the double muon object group, showing exclusive event that contain two muons and different number of jets and b-jets. The numbers on the top of each bin indicate the observed p-value for the agreement of data and simulation for the corresponding event class. The plot on the right shows the invariant mass distribution for the double muon exclusive event class. Measured data are shown as black markers, contributions from SM processes are represented by coloured bars, and the regions enclosed by red dashed lines correspond to the region of interest (contiguous set of bins with lowest p-value) \cite{bib:pas_cms_exo_19_008}.}
\label{fig:music_invmass}       
\end{center}
\end{figure}

 Event classes (final states) are filled based on event content with three types of event classes considered, which are the exclusive event classes, containing events with exactly the specified combination of physics objects, the inclusive event classes, containing events with at least the specified combination of physics objects and inclusive in further physics objects that might be present, and the jet-inclusive event classes, which are inclusive only in the additional number of jets present in the event.
 
 In each event class, a scan is performed on kinematic distributions (invariant mass, sum of transverse momentum, missing transverse momentum) for discrepancies between data and SM to identify regions with significant deviation. A p-value statistic is used to identify deviations and this is corrected for the look-elsewhere-effect.  
If significant deviation beyond expectation is found, it would prompt targeted analysis in the region of interest.
The MUSiC search analysis has been performed using 35.9 $\mathrm{fb^{-1}}$ pp collision data set collected during 2016. 498 exclusive event classes and 571 (530) inclusive (jet-inclusive) event classes with data scanned. Figure~\ref{fig:music_invmass} shows the total event yields in a set of exclusive event classes that contain two muons and different number of jets and b-jets, along with an example of a scan of a kinematic distribution in the dimuon exclusive event class.

No significant deviations beyond expectations were found in the analysed data by the MUSiC algorithm. A wide range of final state topologies have been studied, and agreement has been found between the data and the standard model simulation given the experimental and theoretical uncertainties. This analysis complements dedicated search analyses by significantly expanding the range of final states covered using a model independent approach with the largest data set to date to probe phase space regions beyond the reach of previous general searches.

\section{Summary}
\label{sec:conclusions}
Recent important results on searches for new physics beyond the standard model conducted by the CMS experiment in leptonic final stated have been presented. These searches cover a wide range of potential signal signatures, however no signs of new physics have been found yet. These results represent only a small subset of the large BSM physics search program of the CMS experiment. One can look forward to several more important searches for new physics using the large data set collected by the CMS experiment during the Run-II of the LHC.

\end{document}